\documentclass[useAMS,usenatbib,usegraphicx]{mn2eX}

\usepackage{subfigure}
\usepackage{lscape}

\newcommand{\msun}{\,\mbox{$M_{\odot}$}}

\newcommand{\kms}{\hbox{km s$^{-1}$}}

\newcommand{\degs}{$\degr$}
\newcommand{\chisq}{$\chi^{2}$}

\newcommand{\hd}{\hbox{HD 179949}}
\newcommand{\hdb}{\hbox{HD 179949b}}

\title[\hdb\ 2.1\micron\ contrast ratio]{HD 179949b: A Close Orbiting Extrasolar Giant Planet with a stratosphere?}

\author[J.R.~Barnes et al.]
{J.R.~Barnes$^1$\thanks{E-mail: j.r.barnes@herts.ac.uk} 
 Travis~S.~Barman$^2$,
 H.R.A.~Jones$^1$, 
 C.J.~Leigh$^3$, 
 A.~Collier~Cameron$^4$,
\newauthor 
R.J.~Barber$^5$ and
 D.J.~Pinfield$^1$ \\
$^1$ Centre for Astrophysics Research, University of Hertfordshire, Hertfordshire AL10 9AB. UK. \\ 
$^2$ Lowell Observatory, Planetary Research Centre, 1400 West Mars Hill Road, Flagstaff, AZ 86001. USA \\
$^3$ Astrophysics Research Institute, Liverpool John Moores University, Birkenhead CH41 1LD. UK. \\
$^4$ SUPA, School of Physics and Astronomy, University of St Andrews, Fife KY16 9SS. UK. \\
$^5$ Departments of Physics and Astronomy, University College London, London WC1E 6BT. UK. \\
}

\begin{document}

\date{Accepted 200?. Received 2008}


\maketitle

\label{firstpage}

\begin{abstract}

We have carried out a search for the 2.14 \micron\ spectroscopic signature of the close orbiting extrasolar giant planet, HD 179949b. High cadence time series spectra were obtained with the CRIRES spectrograph at VLT1 on two closely separated nights. Deconvolution yielded spectroscopic profiles with mean S/N ratios of several thousand, enabling the near infrared contrast ratios predicted for the \hd\ system to be achieved.

Recent models have predicted that the hottest planets may exhibit spectral signatures in emission due to the presence of TiO and VO which may be responsible for a temperature inversion high in the atmosphere. We have used our phase dependent orbital model and tomographic techniques to search for the planetary signature under the assumption of an absorption line dominated atmospheric spectrum, where T and V are depleted from the atmospheric model, and an emission line dominated spectrum, where TiO and VO are present.

We do not detect a planet in either case, but the \hbox{$2.120$\ \micron\,-\,$ 2.174$\ $\micron$} wavelength region covered by our observations enables the deepest near infrared limits yet to be placed on the planet/star contrast ratio of any close orbiting extrasolar giant planet system. We are able to rule out the presence of an atmosphere dominated by absorption opacities in the case of HD 179949b at a contrast ratio of $F_p/F_* \sim 1/3350$, with 99 per cent confidence. 

\end{abstract}


\section{Introduction}
\protect\label{section:intro}

Recent models \citep{burrows08cegp} have focused on the role of a parametrized stratospheric absorbing molecular species which leads to abundant molecules such as H$_2$O appearing in emission in certain wavelength bands. Similarly, \cite{fortney08unified} have more specifically investigated the effects of TiO and VO which absorb much of the incoming stellar radiation high in the atmospheres of the hottest CEGPs. \cite{fortney08unified} expect that \hdb\ belongs to this class of hot CEGPs where the absorbing species lead to a temperature inversion and the formation of a stratosphere. It is claimed that such models give better agreement with multi-wavelength mid-infrared Spitzer observations for those CEGPs which are most highly irradiated. Systems such as HD 209458b \citep{burrows07hd209458b,knutson08hd209458b}, HD 149026b and $\upsilon$ And are among those systems which possess atmospheres most consistent with the presence of temperature inversions. 

In the near infrared, the strong \hbox{2.2 \micron}\ bump due to the presence of strong H$_2$O and CO molecular bands for \hbox{$2.2\ \micron\ < \lambda > 2.2\ \micron$} may not be present in systems which exhibit a stratosphere \citep{burrows08cegp}. This possibility was indicated by observations of \hbox{HD 209458b} \citep{richardson03} which failed to detect the \hbox{2.2 \micron}\ bump in the atmosphere of \hbox{HD 209458b}. The weak H$_2$O and CO absorption transitions close to \hbox{2.2 \micron}\ are rather seen in emission while the 2.2 \micron\ $F_p/F_*$ flux ratio is lower and flatter owing to a spectral energy distribution which more closely resembles that of a blackbody.

\begin{table*}
\caption{CRIRES/VLT1 observations of \hd\ for UT 2007 July 26 and August 3. }
\protect\label{tab:journal}
\vspace{5mm}
\begin{center}
\begin{tabular}{lcccccl}
\hline
Object			& UT start of	& UT start of	& Time per	  & Number of		& Number of 	& Comments  \\
			& first frame	& last frame	& exposure [secs] & co-adds per frame	& observations	& 	    \\
\hline
\multicolumn{7}{c}{UT 2007 July 26/27} \\
\hline
HD 179949		& 23:51:46	& 00:06:52	&	25	  & 	4		& 8		&  Main F8V Target	 \\
HD 182645		& 00:18:04	& 00:24:35	&	50	  & 	2		& 4		&  B7IV standard	\\
HD 179949		& 00:39:49	& 03:26:40	&	25	  & 	4		& 76		&  Main F8V Target	 \\
HD 158643		& 00:40:07	& 03:51:41	&	10	  & 	6		& 4		&  A0V Standard	 	\\
HD 179949		& 04:04:28	& 07:54:52	&	25	  & 	4		& 104		&  Main F8V Target	 \\
HD 212581		& 08:09:05	& 08:35:48	&	50	  & 	4		& 8		&  B9.5V Standard	 \\
HD 212301		& 08:48:28	& 09:51:06	&	50	  & 	4		& 16		&  F8V target	 	\\
HD 212581		& 10:05:19	& 10:10:31	&	12	  & 	6		& 4		&  B9.5 Standard 	\\
\hline
\multicolumn{7}{c}{UT 2007 Aug 03} \\
\hline
HD 158643		& 00:18:55	& 00:23:17	&	15	  & 	4		& 4		&  A0V Standard		\\
HD 179949		& 01:08:19	& 02:44:18	&	25	  & 	4		& 44		&  Main F8V Target	\\
HD 158643		& 03:05:54	& 03:17:00	&	10	  & 	6		& 8		&  A0V Standard		\\
HD 179949		& 03:31:11	& 07:22:23	&	25	  & 	4		& 104		&  Main F8V Target	\\
HD 212581		& 07:39:34	& 07:53:31	&	15	  & 	6		& 8		&  B9.5V Standard	\\
HD 212581		& 07:56:14	& 08:36:11	&	20	  & 	6		& 16		&  B9.5V Standard	\\
HD 212301		& 08:49:44	& 09:35:04	&	50	  & 	4		& 12		&  F8V target		\\
HD 212581		& 09:44:38	& 10:24:05	&	20	  & 	6		& 16		&  B9.5V Standard	 \\

\hline
\end{tabular}
\end{center}
\end{table*}

\subsection{HD 179949 and its planet}
\protect\label{section:HD179949}

The presence of a close orbiting planetary system around the \hbox{$M_v = 4.09 \pm 0.04$}\ F8 dwarf star, \hd\ was first reported by \cite{tinney01hd179949}. \cite{eggenberger07} included \hd\ in a search for stellar duplicity but found no companion star, indicating it to be a single star system. Further, a survey of stars known to harbour planets \citep{santos04metallicity} revealed that the [Fe/H] $= 0.22 \pm 0.05$ dex for \hdb\ places it in the most common metallicity band, in a distribution significantly more metal rich than than for stars which are not known to harbour planets. The reported velocity amplitude induced by the planet on its parent star, at \hbox{$K = 101.3 \pm 3.0$ \kms}, with orbital period of $3.093 \pm 0.001$ d\ and orbital radius of $0.045 \pm 0.004$   placed it among the closest orbiting planets. More recently, \cite{butler06catalogue} have published revised system parameters as a result of further monitoring of the system, finding $K = 112.6 \pm 1.8$ \kms, $P = 3.092514 \pm 0.000032$ d and $a = 0.0443 \pm 0.0026$ AU in a low eccentricity ($e = 0.022 \pm 0.015$) orbit. By studying emission in the Ca {\sc ii} H \& K lines, \cite{shkolnik03hd179949} found evidence for perturbation of the stellar magnetic field with a periodicity which coincides with the orbital period of \hdb\ while \cite{wolf04hd179949} found the stellar rotation period of \hd\ to be independent of the planetary orbital period, with $P_{rot} = 7.06549 \pm 0.00061$ d. With no transit of the system reported, a minimum mass of $M sin i = 0.916 \pm 0.076$ \msun\ is found. The time of inferior conjunction is given by an ephemeris of \hbox{HJD = $2451001.510 \pm 0.020$} d \citep{butler06catalogue}. \cite{wittenmyer07} found no evidence for long period objects in the \hd\ system and reported updated system parameters which are consistent with those of \cite{butler06catalogue} at the 1-$\sigma$ level.

\cite{cowan07hotnights} presented mid-infrared Spitzer observations of the \hd\ system which revealed a phase dependent light curve in phase with the planet's orbit and with a relative peak-to-trough amplitude of 0.00141 at \hbox{8 \micron}. This implies that less than 21 per cent of the incident stellar radiation is recirculated to the night side of the planet and contrasts with the other systems in their study, HD 209458 and 51 Peg, which did not reveal photometric variation, suggesting a higher level of redistribution of incident energy. While these observations  were used by \cite{burrows08cegp} to compare model fits, the data carry sufficiently high uncertainties that they enable a number of models with varying degrees of heat redistribution, absorber opacity and inclination to give reasonable fits. A hint that this degeneracy may be partially broken comes from the similarity between \hdb\ and $\upsilon$ And b \citep{harrington06}, both showing only a small shift between the superior conjunction ephemeris and lightcurve maximum. This may indicate that heat is re-radiated without being carried downstream, as suggested by the results of \cite{harrington06,harrington07}, due to the presence of a hot stratosphere which inhibits advection in the lower atmosphere.

In this paper, we present 2.14 \micron\ observations taken with the Cryogenic high-resolution InfraRed \'{E}chelle Spectrograph (CRIRES) at the Very Large Telescope and search for the signature of a planet which exhibits either absorption or emission features. We first describe observations and data extraction procedures (\S \ref{section:obs}) before describing the method (\S \ref{section:analysis}) which involves searching for the faint planetary spectrum in a mean spectrum subtracted timeseries of spectra. A Gaussian matched filter mimics the radial velocity motion and varying strength of the planetary spectrum which is used to search for the best fitting model. The results are presented in \S \ref{section:results} and discussed in \S \ref{section:discussion}.

\section{Observations and Data Reduction}
\protect\label{section:obs}

Observations of \hd\ were made with CRIRES at the Nasmyth focus of the Very Large Telescope, Unit 1 (VLT1, Antu), on 2007 July 26/27 and August 03 and are presented in Table \ref{tab:journal}. Four 1024x1024 InSb Aladdin-3 arrays were used to obtain each spectrum in the wavelength range $\lambda\lambda = 21215\,-\,21740$\ \AA. Observations were made in ABBA nodding sequences to enable subtraction of sky background and lines and to remove the effects of detector defects such as hot pixels. The recorded frames of \hd\ comprised $4 \times 25$ s exposures (see Table \ref{tab:journal}). Observations of early type Standard stars were also made to enable monitoring of changes in telluric line strength throughout each night. For the first and second nights, the seeing varied in the ranges 0.39\arcsec\,-\,0.97\arcsec\ and 0.40\arcsec\,-\,1.4\arcsec\ respectively. The seeing was generally $\sim0.5$\arcsec\ for observations of \hd, while the humidity was very low at 3\,-\,8 per cent and 12\,-\,20 per cent on each night respectively. 

\begin{figure*}
\begin{center}
\includegraphics[height=125mm,angle=0]{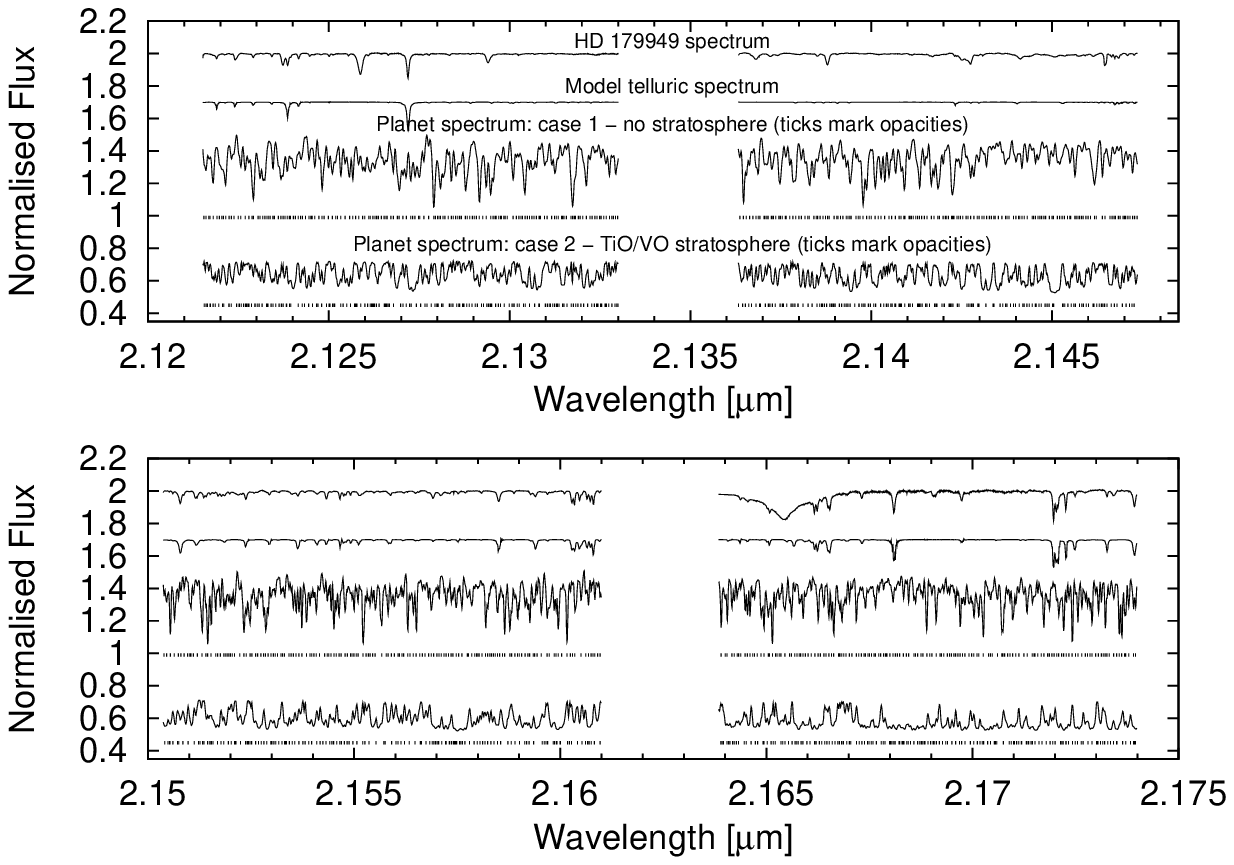}
\end{center}
\caption{CRIRES wavelength coverage of HD 179949. The four segments (left to right and top to bottom) are the spectral regions recorded on detectors 1\,-\,4 respectively. Plotted (top to bottom) in each panel are the mean spectrum for 2007 July 26, model telluric spectrum, model planet spectrum with TiO/VO removed (\hbox{model 1}) and model planet spectrum with TiO/VO stratosphere (\hbox{model 2}). For \hbox{model 2}, the transitions are in emission and are weaker relative to the normalised continuum in the 2.120 \micron\,\,2.174 \micron\ spectral window. The model spectra have been convolved with a Gaussian to match the instrumental resolution. The tick marks represent opacity wavelengths (before convolution) for each planetary model.}
\protect\label{fig:planet_spectrum}
\end{figure*}

\subsection{Data extraction}
\protect\label{section:extract}

The worst cosmic ray events were removed at the pre-extraction stage using the Starlink {\sc figaro} routine {\sc bclean} \citep{shortridge93figaro}. The Aladdin-3 arrays used with CRIRES exhibit an odd/even non-linear sensitivity pattern between adjacent columns (i.e. in the cross-dispersion direction) on detectors 1 and 4 and adjacent rows on detectors 2 and 3 \citep{criresmanual}. The effect is not removed by simple flatfielding or nodding between A and B positions and failure to account for this effect results in extracted spectra which display a saw-toothed pattern on a pixel-to-pixel scale. We implemented the method described by \cite{criresmanual} to account for this effect before further extraction of the science frames was carried out. Each spectrum was then extracted by first subtracting the neighbouring frame in and ABBA sequence to eliminate sky background and detector cosmetic effects. This method means that during extraction of the spectra, it is not strictly necessary to model the sky background. However, we fitted low order polynomials (straight line) across the profile at each dispersion position to account for any slight changes in the sky background from exposure to exposure. Due to the dead space between detectors, it is not possible to record a continuous spectrum with CRIRES over a selected wavelength range.
We therefore carried out the extraction process in the same manner as we would for four individual orders of a cross dispersed \'{e}chelle, with the relevant gain and readout noise for each detector. The typical gain for a detector is 7 electrons/ADU while the r.m.s readout noise is 10 electrons. The spectra were extracted using {\sc echomop}'s implementation of the optimal extraction algorithm developed by \citet{horne86extopt}. {\sc echomop} propagates error information based on photon statistics and readout noise throughout the extraction process. \\

\subsection{Wavelength calibration}
\protect\label{section:wavcalib}

The wavelength range of our observations was chosen as a compromise between minimising the effects of tellurics, maximising the expected number and strength of planetary spectral lines (See \S 4) and maximising the expected planet/star flux ratio. Calibration frames using ThAr lamps were taken during the observations, but only yielded typically one or two lines per order in the observed spectral range. We therefore followed the successful approach used in \citet{barnes07,barnes07b}, where telluric features in the observed spectrum of the standard star were used. A spectrum generated from a HITRAN line list \citep{rothman05hitran} enabled identification of observed features with theoretical wavelengths. A quadratic function was sufficient to describe the non-linearity between pixel position and the wavelength scale, with between 8 and 23 telluric features used for the spectrum on each of the four detectors. A residual r.m.s. of $< 0.042$ times the spectrograph resolution element (a 0.4\arcsec\ slit is equivalent to 0.429 \AA\ at the mean wavelength of the observations ) was obtained resulting in wavelength ranges of \hbox{21214.9\,-\,21330.3 \AA}, \hbox{21363.0\,-\,21473.9 \AA}, \hbox{21503.5\,-\,21610.2 \AA}\ and \hbox{21638.3\,-\,21740.0 \AA} respectively being recorded on each detector.

\section{Data analysis}
\protect\label{section:analysis}

\subsection{Removal of stellar and telluric spectra}
\protect\label{section:removespec}

In order to maximise our chances of detecting a signal we first remove contributions from sources which contaminate the planetary spectrum. We have found that the most effective means of removing the stellar and telluric lines involves subtraction of a scaled and shifted master spectrum \citep{barnes07b}. For each night of observations, a master spectrum is created by cross-correlating and shifting all observed \hd\ spectra to the nearest pixel. The resulting spectrum contains an average of the telluric, stellar and planetary signal with the planetary signal smeared out due to its radial velocity variation with orbital phase. The phase ranges over which we chose to observe \hd\ ensure that any planetary absorption signature will be Doppler shifted from \hbox{-113 \kms}\ to \hbox{+91 \kms}\ during the planet's orbit about \hbox{HD 179949}.

After subtraction of the master spectrum, the remaining principal variation from spectrum to spectrum is seen as a changing telluric line strength throughout the night due to changes in humidity and airmass of the observed target. The telluric lines thus change their strength relative to the stellar lines, and relative to each other. The consequence of the latter effect is that accurate removal of time varying telluric features can only be carried out effectively when individual lines are resolved, as shown by \cite{bailey07}. In \cite{barnes07b}, we fitted the master spectrum to each \hbox{HD 189733} spectrum in turn by taking derivatives of the spectra and using splines to calculate the scale factor at points across the spectra. This process can account for those lines that behave independently over the night and is described in \citet{cameron02upsand} (Appendix A). Care must be taken not to use too few scaling points, which would not adequately account for relative changes in telluric line strength. Conversely, use of too many scaling points could effectively yield a close or perfect match between the individual and master spectrum since the continuum in the master spectrum could be scaled to mimic a weak absorption (e.g. planet) feature. This would result in attenuation or removal of a potential planetary signal. We therefore do not include more scaling points than twice the effective spectral resolution. With these limitations, the effectiveness of our spectrum removal procedure is dependent on spectral resolution, and at the resolution of R $\sim15,000$ \citep{barnes07b} significant residuals inevitably remained. In contrast, the present CRIRES data have R $\sim50,000$, which enable more accurate telluric subtraction. 

Some tellurics nevertheless still remain at the 2-$\sigma$ level (\S \ref{section:results}) after subtraction of the scaled master spectrum. We therefore reduced the strength of remaining systematics with principal components analysis of the spectral timeseries using the method first described in \cite{cameron02upsand} (Appendix B) and discussed in both \citet{barnes07} and \citet{barnes07b}. The data at fixed wavelength positions are decomposed into a series of eigenvectors which the main orthogonal trends in the timeseries of observations. Removal of the first four principal components ensure that the remaining systematics due to time varying telluric residuals are subtracted while not significantly attenuating a moving planetary signal. At the high S/N ratios (see \S \ref{section:atmos}) of the observed spectra, complete reduction of residuals to levels close to 1-$\sigma$ is not possible.

\subsection{Deconvolution and Atmospheric Model}
\protect\label{section:atmos}

\begin{figure}
\centering
\begin{center}
\includegraphics[height=75mm,angle=270]{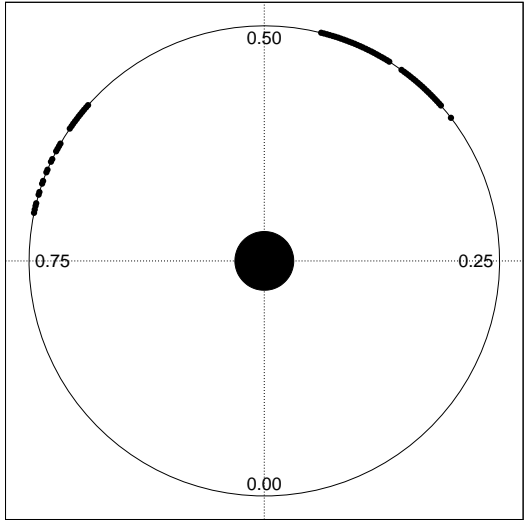}
\end{center}
\caption{Orbital phase diagram for HD 179949b. Phase 0.00 corresponds to the time of inferior conjunction. The large filled circle at centre represents HD 179949 while the small filled circles represent the phases of observations.}
\protect\label{fig:orbit}
\end{figure}

To enable absorption or emission features to be detected, we deconvolve our observed spectra using a model of the planetary atmosphere. In effect, a least squares deconvolution \citep{donati97zdi} is applied to each observed spectrum by using the wavelength positions and line depths of all opacities in the wavelength range of interest \citep{cameron99tauboo,barnes07,barnes07b}. For \hbox{HD 179949b}, we model an atmosphere with solar metallicity for a temperature, \hbox{T = 1250 K} and surface gravity, \hbox{log $g = 1.33$ ms$^{-1}$}. The effective wavelength of the single mean profile is determined by the distribution of line opacities, opacity strengths and number of counts at each wavelength \citep{barnes98aper}. The mean weighted effective wavelengths of the deconvolved profiles are $\lambda_c = 2.1434$ \micron\ and $\lambda_c = 2.1428$ \micron\ for \hbox{model 1} and \hbox{model 2} respectively. The similarity is a reflection of the uniformity in strength and distribution of opacities in the two scenarios. 

We have discussed the sensitivity of the model to mismatches between in the observed and model spectra in \cite{barnes07,barnes07b}. For detection of absorption/emission features it is necessary that a model with opacities which are as up to date as possible are used. Our models are generated using the \cite{barber06water} water line list, H$_{2}$O being the dominant species in the wavelength region of our observations. We have discussed the role of absorbing species (\S \ref{section:intro}) in the upper atmospheres of those planets which absorb the highest amount of incident solar radiation. For cooler atmospheres, Ti and V are believed to rain out of the upper atmosphere, thereby giving no contribution to emergent spectra. In the case of the hottest CEGPs however, T and V are present as oxides, namely TiO and VO, which absorb incident radiation high in the atmosphere leading to heating and formation of a temperature inversion. It is claimed by \cite{burrows08cegp}, \cite{burrows07hd209458b} and \cite{knutson08hd209458b} (HD 209458 b) that multi-wavelength mid-infrared observations are in better agreement with this scenario. \cite{fortney08unified} have divided cooler and hotter planets into two regimes which they call pL and pM class respectively. Both \cite{burrows08cegp} and \cite{fortney08unified} discuss whether \hd\ belongs to the former or latter of these classes, but given the relatively early spectral F8-F9V spectral type of \hbox{HD 179949} itself, it seems likely that \hd\ belongs to the pM class.

Building upon the work of \cite{barman05}, model spectra are generated for the two cases (1) where TiO and VO are removed from the atmosphere, leading to dominance of strong bands of H$_{2}$O absorption (in the region of interest) and (2) where TiO and VO lead to formation of a stratosphere, resulting in formation of weak lines in emission (hereafter \hbox{model 1} and \hbox{model 2}). The form of the spectrum in \hbox{model 2} is much more blackbody-like than in \hbox{model 1} and results in an overall lower value for the flux ratio in the wavelength region of our observations. For a fuller description of the model opacities and setup see \citet{ferg05}, \citet{barman01} and \citet{barman05}. The resulting models are expected to be very similar to those of \cite{burrows08cegp} and \cite{fortney08unified} described above. Fig. \ref{fig:planet_spectrum} shows the mean normalised spectrum from July 26, the model normalised telluric spectrum \citep{rothman05hitran} and the normalised planetary spectra for models 1 and 2.

After co-addition of ABBA sequences, a total of 46 spectra were observed on July 26 and 27 spectra were observed on August 2. The mean S/N ratio, measured from residual ABBA spectra after removal of the template and principal components, was \hbox{$959 \pm 57$} and \hbox{$893 \pm 73$} for the night of 26 July and 02 August respectively with a combined S/N ratio of \hbox{$935 \pm 71$}. Deconvolution using \hbox{model 1} yielded S/N = $8263 \pm  721$ while \hbox{model 2} yielded $2589 \pm 252$, indicating effective gains of 8.84 and 2.77 respectively. The lower S/N achieved through use of \hbox{model 2} is a consequence of the weaker emission lines (see Fig. \ref{fig:planet_spectrum}).


\subsection{Matched filter}
\protect\label{section:matchedfilter}

We aim to isolate the phase dependent planetary absorption signature from the total light of the \hd\ system. Unlike the stellar spectrum, the planetary spectrum is Doppler shifted due to its orbit and is expected to exhibit a phase variation in strength due to irradiation of its inner face \citep{barman05}. The time-dependent changes in Doppler shift and brightness are modelled by using a Gaussian matched filter which has been described in both \citet{barnes07} and \citet{barnes07b}. The method is a adapted from the reflected light studies which are detailed in \hbox{Appendix D} of \citet{cameron02upsand}. By modelling the variation in Doppler shifted velocity and planet/star flux ratio ($F_p/F_*$), we are able, in the presence of a strong detection to determine both the maximum planet/star flux ratio ($\epsilon_0 = F_p/F_*$) and the velocity amplitude of the planet, $K_p$. Fig. \ref{fig:orbit} shows the phases at which observations were made. Phases close to $\phi = 0.5$ are favoured because the hot inner face is presented to the observer, resulting in the highest planet/star flux ratios.

\subsection{Signal calibration and orbital inclination}
\protect\label{section:signalcalib}

The matched filter is calibrated by means of a simulated fake planetary signal with known $\epsilon_{0}$ and $K_p$. The fake signal is added to the extracted spectra before further analysis, which involves removal of the mean stellar spectrum and residual noise patterns. As \hd\ is not a transiting system, the axial inclination of \hbox{HD 179949} is unknown. However, since Rossiter-McLaughlin effect measurements for a number of systems, including HD 209458 \citep{bundy00hd209458, queloz00hd209458, winn05, wittenmyer05} and HD 189733 \citep{winn06}, have shown that the orbital plane of the planet closely matches that of the axial inclination of the star, we carry out Monte-Carlo simulations to determine the most probable velocity amplitude, $K_p$, and the orbital inclination, $i$, of \hdb. Using the empirical orbital parameters discussed in section  \S \ref{section:intro}, we find most probable values of $K_p = 115$ \kms\ and $i = 47.6$\degs. A matched filter 2D search as described above (\S \ref{section:matchedfilter}) for the best fitting parameters is scaled to ensure the the planet is recovered at the simulated value of $\epsilon_0$ for the fake planet.

The significance of the result is assessed using bootstrap statistical procedures. The data are re-ordered in a way which randomises the observation phases while preserving the effects of any correlated systematic errors \citep{cameron02upsand}. Essentially, the order of the observations is randomised in a set of 3000 trials, thereby scrambling any true planetary signal while enabling the data to retain the ability to produce spurious detections through the chance alignment of systematic errors. The least squares estimate of log$_{10}\,\epsilon_{0}$ and associated \chisq\ as a function of $K_p$ enable us to plot 68.4, 95.4, 99.0 and 99.9 per cent bootstrap limits on the strength of the planetary signal.

\section{Results}
\protect\label{section:results}

\begin{figure*}
 \begin{center}
   \begin{tabular}{cc}

      \vspace{10mm} \\
      \includegraphics[width=70mm,bbllx=113,bblly=114,bburx=397,bbury=397,angle=0]{barnes_hd179949_fig3.ps} \hspace {5mm} &
      \hspace {5mm}
      \includegraphics[width=70mm,bbllx=113,bblly=114,bburx=397,bbury=397,angle=0]{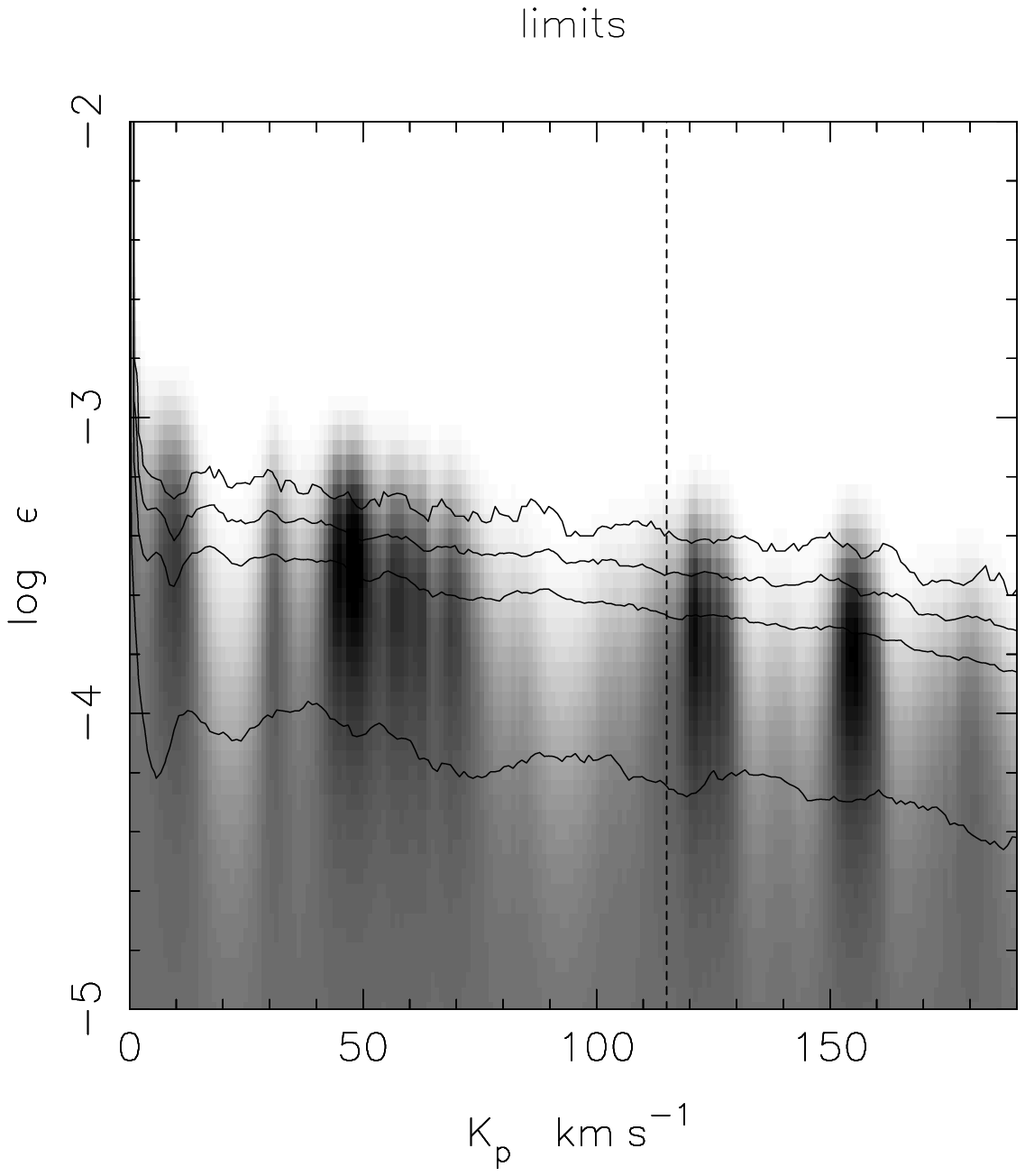} \\

      \vspace{20mm} \\
      \includegraphics[width=70mm,bbllx=113,bblly=114,bburx=397,bbury=397,angle=0]{barnes_hd179949_fig5.ps} \hspace {5mm} &
      \hspace {5mm}
      \includegraphics[width=70mm,bbllx=113,bblly=114,bburx=397,bbury=397,angle=0]{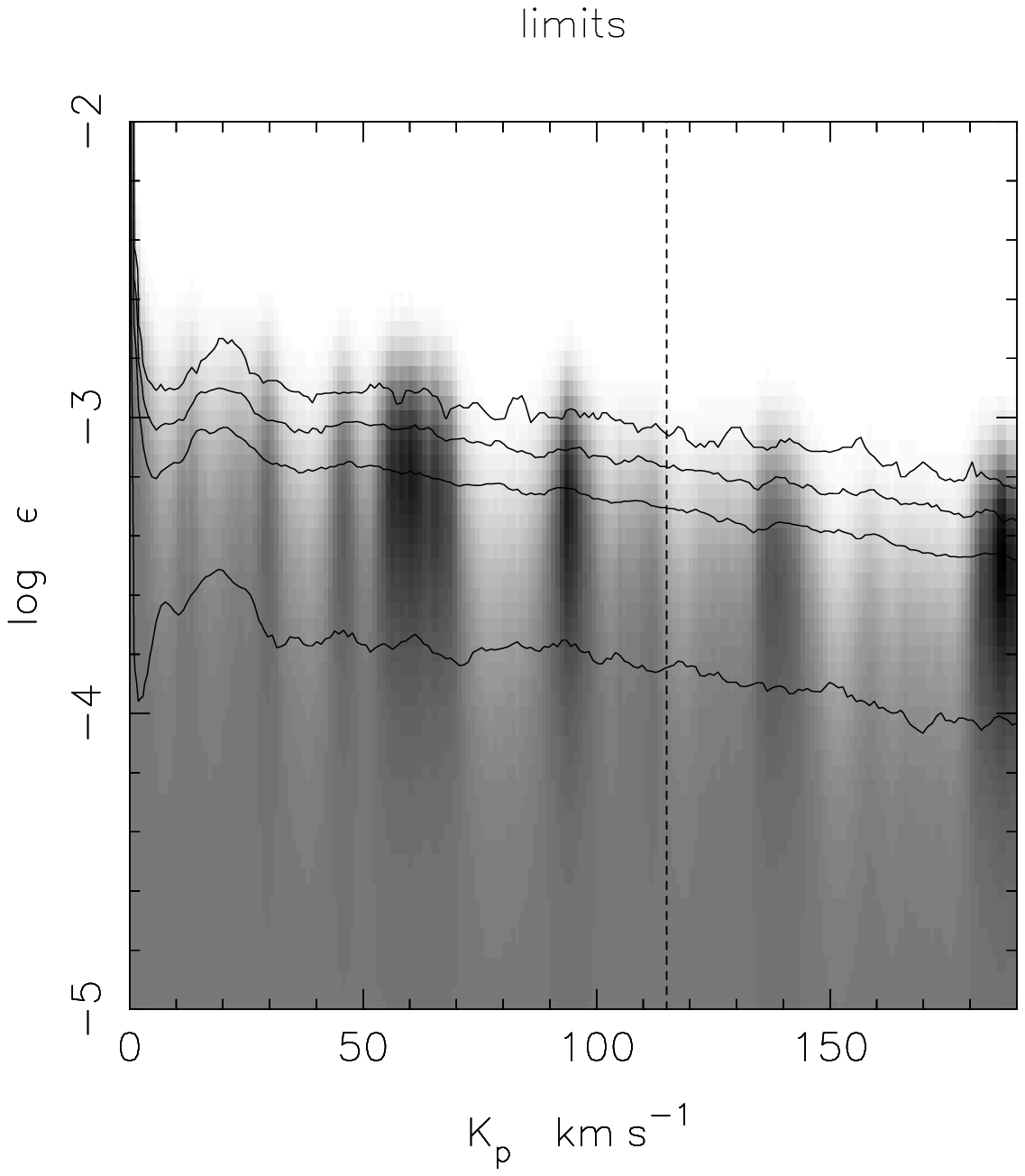} \\

      \vspace{10mm} \\
   \end{tabular}
 \end{center}

\caption{Left: Phased timeseries of the deconvolved residual spectra of \hd. The dashed sinusoidal curve represents the {\em expected} motion of a planetary signal based on the most probable velocity amplitude, \hbox{{\em \^K}$_p = 115$}~\kms. For plotting purposes, all deconvolved profiles have been normalised to the same noise level to optimise the visible information content of the plot. Right: Relative probability \chisq\ map of planet-star flux ratio $log_{10}\,\epsilon_{0}$ vs $K_p$. The first 4 principal components have been subtracted from the timeseries spectra. The greyscale in the probability maps represents the probability relative to the best fitting model in the range white for 0 to black \hbox{for 1}. Numerous enhancements in \chisq\ result from low level systematic residuals (see main text) which can not effectively be removed. Plotted are the 68.3, 95.4, 99 and 99.9 per cent (bottom to top) confidence levels for a planetary signal detection. No planetary signal is detected at the most probable velocity amplitude, \hbox{{\em \^K$_p$}}, which is represented by the vertical dashed line.}
\protect\label{fig:results}
\end{figure*}

We present the results of our search for the 2.143 \micron\ high resolution spectroscopic opacity signatures of \hdb\ in Fig. \ref{fig:results}. The timeseries deconvolved spectra spanning both nights of observations are shown in Fig. \ref{fig:results} for \hbox{model 1} (top left) and \hbox{model 2} (bottom left). In the former scenario, since the opacities are expected to appear in absorption, a planetary signature would be seen as a dark trail while in the latter scenario, a planetary signature would be seen as a light trail relative to the mean background. In each case, the dashed line represents the velocity path of a planetary signature at the most probable velocity amplitude, \hbox{{\em \^K$_p = 115$} \kms}, and does not represent a detection.

The dark regions in the \chisq\ plots of $log_{10}\,\epsilon_{0}$ vs $K_p$ represent local enhancements in \chisq\ for particular combinations of parameters. These features are all present within the 68.3, 95.4 per cent confidence (1-2 $\sigma$) intervals and are due to systematic features which remain in the residual spectra. It is possible to rule out a number of features where the false alarm probability would be high. An estimate of the upper limit to the velocity amplitude of $K_p({\rm max}) = 155.8 \pm 9.1$ \kms\ is determined from the orbital period and semi-amplitude. Planets detected at velocities much in excess of this limit can thus be ruled out. Similarly, a planet with low orbital inclination would be expected to yield a much lower planet/star flux ratio owing to the heated inner face of the planet appearing foreshortened to the observer.

For \hbox{model 1}, we are unable to detect a planetary signal at the most probable velocity amplitude at a level of $log_{10}\,\epsilon_{0} = -3.67$ with 95.4 per cent confidence (2-$\sigma$). This equivalent to $F_p/F_* \sim 1/4600$. Similarly, for \hbox{model 2}, we do not detect a planetary signal with $log_{10}\,\epsilon_{0} = -3.30$ ($F_p/F_* \sim 1/2000$) with 95.4 per cent confidence (2-$\sigma$). The implications and significance of these results are discussed in \S \ref{section:discussion}.

The epoch of inferior conjunction and orbital period (phase, $\phi = 0.0$) of \hd\ carry uncertainties which may result in significant drift of the phasing by the time of observations. A drift of up to $\pm$\ 1hr 18 min or $\phi = \pm 0.0175$ could have accumulated from the ephemeris up to the mid-epoch of the present observations. Since such a drift in observations may be responsible for our inability to detect the planetary signal, we investigated shifting the time of inferior conjunction by $\pm 15$ mins intervals in this range. No clear candidate signature was however detected at any of the shifted ephemerides.

\section{Discussion}
\protect\label{section:discussion}

\begin{figure}
\begin{center}
\includegraphics[height=62mm,angle=0]{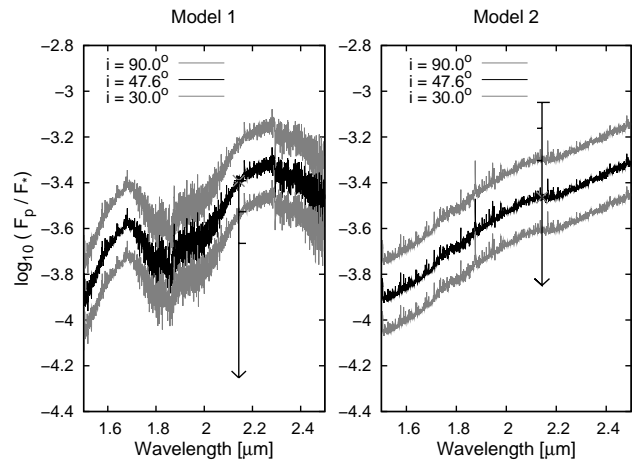}
\end{center}
\caption{Model and 2.14 \micron\ planet/Star flux ratios for HD 179949b. Shown are models for \hbox{model 1}, absorption (left) and \hbox{model 2}, TiO/VO stratosphere, emission spectrum (right). Some apparent emission features are due to {\em stellar absorption} lines and result from dividing the planet spectrum by the stellar spectrum. The black crosses indicate the mean model flux over the wavelength of observations for the most probable inclination. Upper limits corresponding to the levels in Fig. \ref{fig:results} (99.9, 99 and 95.4 per cent) are marked by horizontal bars (from top to bottom) for our analysis using each model. The 99.9 per cent observational upper limit is coincident with the mean flux level for the \hbox{model 1} model. The width of the top horizontal bar in each case (99.9 percent confidence) represents the wavelength range of the CRIRES spectroscopic observations.  For \hbox{model 1}, we can marginally exclude the model with 99.9 per cent confidence at the most probable inclination of 47.6\degs. For \hbox{model 2} we are unable to exclude the model since the 95.4 per cent confidence level (2-s$\sigma$) is above the model flux ratio and artifacts are present between 2-s$\sigma$ and 1-s$\sigma$ (68.3 per cent confidence marked by the arrow).}
\protect\label{fig:planet_ratios}
\end{figure}

The significance of our analysis is presented in Fig. \ref{fig:planet_ratios} which shows the model and observed planet/star flux ratios. Fig. \ref{fig:planet_ratios} (left) indicates the 99.9, 99, 95.4 and 68.3 per cent upper detection limits for an atmosphere which contains no high altitude absorbing species which lead to a temperature inversion. In this model, only absorption lines are present and our analysis is able to reject this scenario with 99 per cent confidence at a level of $log_{10}\,\epsilon_{0} = -3.53$ ($F_p/F_* \sim 1/3350$) for a planet with the most probable velocity amplitude. The \hbox{model 2} scenario, where TiO and VO are able to absorb incoming radiation high in the atmosphere and form a temperature inversion, leads to a spectrum resembling a blackbody with weak emission lines superimposed (Fig. \ref{fig:planet_ratios}, right). Since the lines are weak, we are unable to reliably reject this model with our analysis since the 95.4 per cent (2-$\sigma$)  upper limit of $log_{10}\,\epsilon_{0} = -3.30$ ($F_p/F_* \sim 1/2000$) at the most probable velocity amplitude is higher than the model contrast ratio of $log_{10}\,\epsilon_{0} = -3.46$ ($F_p/F_* \sim 1/2900$) at the mean wavelength of the observations.

Without a clear detection of the signal, it is difficult to constrain the model predictions further since the flux ratio is dependent on the as yet unknown orbital inclination of \hdb. The prior distribution for $K_p$ indicates that the relative chance of detecting the planet with $K_p({\rm max})$ (i.e. $i = 90$\degs) is $\sim 9.7$ per cent of a detection at \hbox{{\em \^K$_p$}}. Similarly, an orbit with i.e. $i = 30$\degs\ carries only a 4.7 per cent relative chance. These extremes are represented in Fig. \ref{fig:planet_ratios} by the grey spectra which have simply been scaled by the flux expected from the planet at $i = 30$\degs\ and $i = 90$\degs. Even with $i = 30$\degs, we can rule out the presence of a planet with a \hbox{model 1} atmosphere with 99 per cent (marginal) confidence. Conversely, for \hbox{model 2}, were $i = 90$\degs, we would marginally detect a planet with 99 per cent confidence. There is no evidence for eclipse events in \hdb\ however.

In \cite{barnes07,barnes07b}, we investigated the effect of model-observation line strength mismatch. However, the transitions of the strongest H$_2$O lines which contribute to the model spectrum \citep{barber06water}, and hence the deconvolved spectra, are known experimentally, with significant uncertainties on only the very weakest lines.  If we can rule out wavelength and oscillator strength uncertainties in the input H$_2$O transitions as sources of mismatch between model and observation, uncertainties in line strength may arise from the planetary temperature structure which may be incorrect due to the assumptions of hemispheric-wide hydrostatic equilibrium.

Until a clear detection of a planetary signature is made at high resolution, a quantitative {\em empirical} model-observation mismatch constraint will not be possible. Nevertheless, successful detection of H$_2$O at optical \citep{barman07water} wavelengths, and of H$_2$O and CH$_4$ via transmission spectroscopy at mid-infrared wavelengths \citep{tinetti07nature,swain08methane}, have relied upon models \citep{tinetti07transmission} which contain reliable molecular opacities such as the H$_2$O transitions of \cite{barber06water}. The mid-infrared observations of the spectral energy distribution of HD 189733b by \cite{tinetti07nature} and \cite{swain08methane} rely on relatively broad spectral regions where the match between observation and model appears to be very close. While transmission spectroscopy is sensitive to the planetary atmospheric signature at the terminator, the same opacities are included in the model we have used for our dayside average spectrum, and have been used by \cite{barman08hd189733b}, to infer the presence of H$_2$O and CO in the atmosphere of \hbox{HD 189733b}.

Our inability to detect H$_2$O (the only species in the wavelength interval of our observations) in the atmosphere of HD 179949b, can be explained by the expected \hbox{model 2} scenario and lack of required sensitivity in our data. On the other hand, for \hbox{HD 189733b}, we rejected the expected \hbox{model 1} scenario \citep{barnes07b} using the same technique as presented in this paper. This may be explained by a lower than expected contrast ratio resulting from efficient flux redistribution at the deeper layers probed by shorter wavelengths \citep{barman08hd189733b}. This could also apply to our rejection of the \hbox{model 1} scenario for \hdb. However, given the earlier spectral type of \hd\ when compared with \hbox{HD 189733}, and subsequent higher irradiation of \hdb, as already discussed, the \hbox{model 2} scenario seems more appropriate \citep{fortney08unified,burrows08cegp}. It is worth noting that the deeper atmospheric layers probed by the shorter wavelength observations in the model 1 scenario may be blocked by the presence of the high altitude absorber(s) in the \hbox{model 2} scenario. Hence a non-detection of \hdb\ with higher S/N ratio observations could not be explained in the same way as for \hbox{HD 189733b}.

\section{Conclusion}
\protect\label{section:conclusion}

We have applied a planetary atmospheric model with two physically different conditions in an attempt to detect the planetary signature of \hdb. We are able to reject the scenario in which Ti and V are depleted from the atmosphere of \hdb\ in agreement with the predictions of recent models. We have also assessed our ability to test the models in which TiO and VO are present in the upper atmosphere, resulting in a temperature inversion and lines appearing in emission. We are however unable to detect a planet using this model, or to reject the model since our data do not achieve the necessary S/N ratios required. The 99 per cent confidence limit is 2 times higher than the expected flux ratio at the most probable velocity amplitude of the planet and would thus necessitate a further six nights of observation before this level of sensitivity could be reached.

In order to reliably test the most recent models which include a stratosphere and weak emission, cross dispersed spectrographs which offer greater wavelength coverage will afford higher S/N gain and thus improve our chances of detection. Equally, study of transiting systems reduce the degeneracy considerably in any analysis since the inclination is known and will be close, or equal to 90\degs. The greater maximum brightness of a transiting CEGP system will increase our chances of detecting and characterising the planet.

\section{Acknowledgments}
JRB was supported by a STFC funded research grant during the course of this work. TB acknowledges support from NASA's Origins of Solar System program and the NASA Advanced Supercomputing facility, and LP from NSF grant 04-44017. The data presented herein were obtained at UT 1 of the Very Large Telescope array of the European Southern Observatory. JB wished to thank

\bibliographystyle{mn2e}
\bibliography{iau_journals,planets,master,ownrefs}

\end{document}